# Probing dense QCD matter: Muon measurements with the CBM experiment at FAIR

A. Senger[1,*] and P. Senger[1,2] for the CBM collaboration

[1] Facility for Antiproton and Ion Research, Darmstadt, Germany

[2] National Research Nuclear University MEPhI, Moscow, Russia

*Correspondence: a.senger@gsi.de

**Abstract**

The Compressed Baryonic Matter (CBM) experiment at the future Facility for Antiproton and Ion Research (FAIR) in Darmstadt is designed to investigate the properties of high-density QCD matter with multi-differential measurements of hadrons and leptons, including rare probes like multi-strange anti-hyperons and charmed particles. The research program covers the study of the high-density equation-of-state of nuclear matter, and the exploration of the QCD phase diagram at large baryon-chemical potentials, including the search for quark matter and the critical endpoint of a hypothetical 1$^{st}$ order phase transition. The CBM setup comprises detector systems for the identification of charged hadrons, electrons, and muons, for the determination of collision centrality and the orientation of the reaction plane, and a free-streaming data read-out and acquisition system, which allows online reconstruction and selection of events up to reaction rates of 10 MHz. In this article, emphasis is placed on the measurement of muon pairs in Au-Au collisions at FAIR beam energies, which are unique probes to determine the temperature of the fireball, and, hence, to search for a caloric curve of QCD matter. Simultaneously, the subthreshold production of charmonium can be studied via its dimuon decay, in order to shed light on the microscopic structure of QCD matter at high baryon densities. The CBM setup with focus on dimuon measurements and results of the corresponding physics performance studies will be presented.

1. Introduction

The Compressed Baryonic Matter (CBM) experiment at the future Facility for Antiproton and Ion Research (FAIR) is designed to study the properties of QCD matter at very high baryon densities in heavy-ion collisions with kinetic beam energies between 2A and 11A GeV. The research program covers the high-density equation-of-state (EOS) of nuclear matter, and the exploration of the QCD phase diagram at high densities, including the search for structures like the critical endpoint of a 1$^{st}$ order phase transition. As diagnostic probes, CBM will measure a broad variety of observables, including the collective flow of identified particles, event-by-event fluctuations of multiplicity distributions of conserved quantities, yields and phase-space distributions of multi-strange (anti-) hyperons, hypernuclei, charmed particles, and lepton pairs.

Electrons and muons born in the hot and dense fireball of a heavy-ion collision do not suffer from strong absorption and rescattering effects, and carry undisturbed information to the detector system. Lepton pairs are either decay products of vector mesons, or, like photons, are thermally radiated away from the hot matter. At invariant masses up to about 1 GeV/c$^2$, the spectrum is dominated by decays of η, ω, ρ, and φ mesons. The measurement of their decay leptons, in particular from the short-lived ρ mesons, provides information on their in-medium mass modifications due to chiral symmetry restoration [1]. Above 1 GeV/c$^2$, where the contributions from vector meson decay are strongly reduced, the dilepton invariant mass spectrum reflects the average temperature of the emitting source, integrated over the entire collision history. The source temperature can be directly extracted from the slope of the dilepton invariant mass spectrum, as it is not affected by the radial flow. Such a measurement has been performed by the NA60 collaboration at the CERN-SPS, who extracted an average source temperature of 205 ± 12 MeV from the μ+μ- invariant mass spectrum measured in In+In collisions at a beam kinetic energy of 158A GeV [2]. Recently, the HADES collaboration managed to extract a source temperature of 72 ± 2 MeV from the low-mass region of the di-electron invariant mass spectrum measured in Au+Au collisions at a beam kinetic energy of 1.25A GeV, by subtraction of the known contribution from vector meson decays [3].

The NA60 and HADES data points are included in figure 1, which depicts the temperature of the matter produced in heavy-ion collisions as function of collision energy [4]. The dashed magenta line represents the average fireball temperature, i.e. the slope of the dielectron invariant mass spectrum between 1 and 2 GeV/c$^2$. Up to a collision energy of $\sqrt{s_{NN}}$ = 6 GeV, the spectrum was simulated by applying a coarse-graining method to an hadronic UrQMD transport model calculation [5], whereas the high-energy part ($\sqrt{s_{NN}}$ > 6 GeV) of $T_{slope}$ as well as $T_{initial}$ was calculated with a fireball model [6]. The blue solid line illustrates the parameterization of the freeze-out temperature, i.e. at a late stage of the collision, when the particles seize to interact inelastically [7]. Recent lattice QCD calculations found a relatively low chiral phase transition temperature of $T^0_c$ = 132+3-6 for a baryon chemical potential $\mu_B$ = 0 in the chiral limit [8]. According to this calculation, the critical endpoint of a 1st order phase transition for physical quark masses and finite $\mu_B$ will be located at an even lower temperature, if it exists at all. This upper temperature limit for the critical endpoint is indicated by the green dashed area in figure 1.

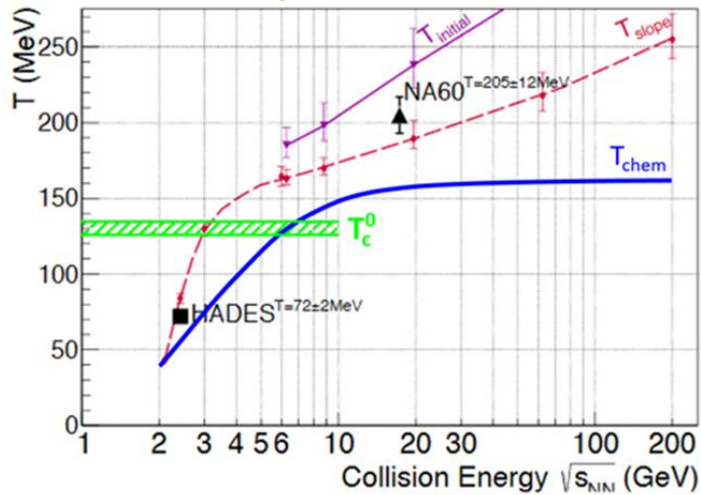

Fig. 1: Fireball temperature as function of collision energy [4]. The magenta dashed line is derived from calculations of the dilepton invariant mass distribution in the range above 1 GeV/c$^2$ ($T_{slope}$) using a coarse-graining method applied to a transport model for energies up to $\sqrt{s_{NN}}$ = 6 GeV [5], and with a fireball model for energies above $\sqrt{s_{NN}}$ = 6 GeV [6]. The purple line, representing the initial temperature, was also calculated with the fireball model. The NA60 [2] and HADES [3] data points are extracted from measured dilepton invariant mass spectra (see text). The blue solid line represents the chemical freeze-out curve [7], and the green dashed area is the upper bound for the temperature of the critical endpoint calculated by lattice QCD [8].

A unique method, to identify a possible 1st order phase transition, and to discover or to rule out the existence of a critical endpoint, is the measurement of the fireball temperature via lepton pairs in heavy-ion collisions at different beam energies. If the temperature scan results in a caloric curve, both phase coexistence and its endpoint is determined. Moreover, according to the lattice QCD calculation the onset of deconfinement and the critical endpoint should be located in the FAIR energy range. Therefore, a fundamental aspect of the CBM research program will be the precise study of both dielectron and dimuon emission in heavy-ion collisions at collision energies between $\sqrt{s_{NN}}$ = 2.5 and 5 GeV, where a 1st order phase transition should happen, if it exists. Likewise, the measurement of a smooth temperature curve, which would rule out the existence of a critical endpoint, would be a very important result.

A smooth excitation function of the fireball temperature in the FAIR energy range, where the net-baryon density varies from 2 to about 8 $\rho_0$ depending on the beam energy [9], might either indicate a continuous crossover from hadronic to quark matter, or the absence of a phase transition. The measurement of charmonium might shed light on the microscopic degrees-of-freedom at these densities, and help to disentangle the two scenarios. As the thresholds for charmonium production in nucleon-nucleon collisions are slightly above the gold-beam energy at FAIR/SIS100, the identification of J/ψ mesons in Au+Au collisions with the CBM experiment would strongly favor a hadronic scenario, as suggested by UrQMD calculations [10]. Within this model, charmed particles are produced via the decay of heavy N* resonances, in processes like N*→J/ψ+N+N and N*→$\Lambda_c$+anti-D, assuming that the N* have been excited by sequential hadron-hadron collisions. A big experimental advantage is, that both the J/ψ mesons and the radiation from the fireball can be measured simultaneously by dilepton pairs in the same CBM setup.

Dilepton measurements in heavy-ion collisions are extremely challenging because of the large combinatorial background, which has different sources for electrons and muons. Therefore, the CBM experiment comprises detector systems for the alternative identification of $e^+e^-$ and $\mu^+\mu^-$ pairs, in order to reduce the systematic uncertainties. In the following, the CBM experimental setup will be briefly described. The focus will be on the layout of muon detection system, on the methods of track reconstruction and particle identification, and, finally, on physics performance studies, in order to demonstrate the capabilities of the detector system with respect to the physics program sketched above.

2. The CBM experimental setup

A sketch of the CBM detector system is shown in figure 2. The setup comprises two silicon detector systems, located in the large-aperture gap of a superconducting dipole magnet, for the reconstruction of up to 700 charged particle tracks per central Au+Au collision. The first detector the Micro-Vertex Detector (MVD) consisting of four layers Monolithic Active Pixel Sensors (MAPS). The second Silicon Tracking System (STS) comprises eight stations with in total 900 double-sided micro-strip silicon sensors. For electron measurements, a Ring Imaging Cherenkov (RICH) detector and a Transition Radiation Detector (TRD) are located downstream the magnet. While the RICH suppresses pions up to momenta of about 7 GeV/c, the TRD mainly reduces the yield of pions with larger momenta. For muon measurements, the RICH will be replaced by a muon chamber (MuCh) system, comprising hadron absorbers and tracking chambers. In this case, which is illustrated in figure 1, the TRD provides muon tracking behind the last hadron absorber. Particle identification is performed by Time-of-Flight (TOF), which consists of Multi-gap Resistive Plate Chambers, and is positioned downstream the TRD. The last device, located behind the TOF wall, is a hadron calorimeter consisting of 44 modules, the so called Projectile Spectator Detector (PSD), which provides the determination of collision centrality and reaction plane. The CBM detector systems are described in more detail in [11], together with the free-running high-speed data acquisition system.

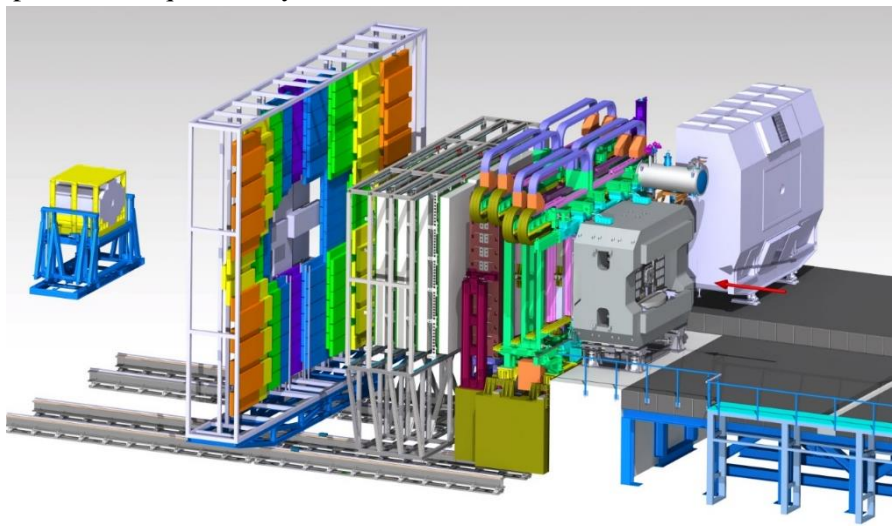

Fig. 2: The CBM experimental setup at FAIR (see text). The beam enters from the right side.

3.    The Muon Detection System of CBM

Up to date, no muon measurements have been performed in heavy-ion collisions below the top SPS energy of 158A GeV. In the much lower FAIR energy range, i.e. at beam energies between 2 - 11A GeV, the challenge is to identify muons with low momenta in an environment of high particle multiplicities. For example, in a central Au+Au collision at a beam energy of 8A GeV, less than 20 vector mesons (mostly η, ρ, ω) are produced, which decay into $\mu^+\mu^-$ pairs with a branching ratio of about $10^{-4}$. Hence, the experimental task is to find one created $\mu^+\mu^-$ pair in 500 collisions, each collision producing about 700 charged hadrons, which should be suppressed by absorbers. The surviving hadrons contribute to the combinatorial background when reconstructing $\mu^+\mu^-$ pairs. As the momentum of the muons depend on beam energy and on the invariant mass of the muon pair, the number of hadron absorbers in the CBM muon detector system will be chosen accordingly. The three different configurations are sketched in figure 3. The minimum version is depicted in the left panel of figure 3. It consists of a 60 cm thick absorber made of carbon and concrete followed by three tracking stations, a 20 cm iron absorber followed be 3 tracking stations, and another 20 cm iron absorber followed by the TRD as tracking station. This configuration will be used for beam energies from 2A to 4A GeV, and for studies of vector mesons, i.e. for dimuon invariant masses up to about 1 GeV/$c^2$. In order to study at these energies the intermediate dimuon invariant mass region up to about 3 GeV/$c^2$, another 30 cm iron absorber followed by three tracking stations will be added, as shown in the center panel. This version will be also used for beam energies above 4A GeV, but only for the low invariant mass region. In order to measure the high dimuon invariant masses at higher beam energies, another 1 m thick iron absorber has to be added, in order to reduce the combinatorial background. This configuration is depicted in the right panel of figure 3. In all three configurations, the TRD serves as a tracker, and the TOF detector is used for additional hadron suppression and muon identification.

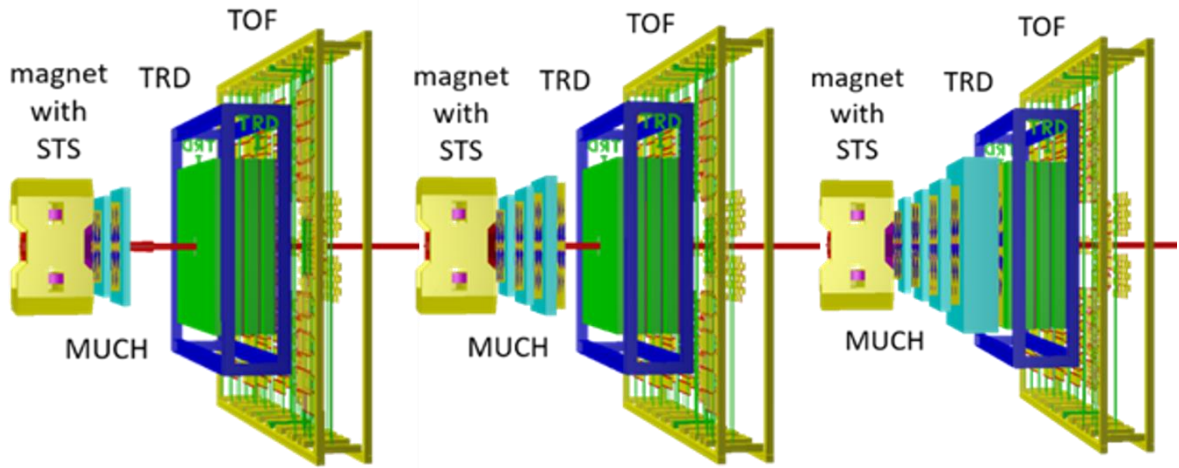

Fig. 3: Three configurations of the CBM-muon setup, which comprises the dipole magnet with the Silicon Tracking System inside, the Muon Chamber (MuCh) system, the Transition Radiation Detector (TRD) as additional tracking system, and the Time-of-Flight (TOF) detector. Left panel: MuCh with 3 absorber layers and 2 tracking stations Center panel: 4 absorber layers and 4 tracking stations. Right panel: 5 absorber layers and 4 tracking stations (see text).

The layout and technology of the much tracking detectors depend on hit density and rate. The first two tracking stations each consist of 3 layers with 16 respective 20 trapezoidal shaped Gas Electron Multiplier (GEM) modules. The readout pads varying in size from about 10 mm$^2$ at small emission angles up to 440 mm$^2$ at large angles, which is matched to particle hit densities up to 500 kHz/cm$^2$. The 3$^{rd}$ and 4$^{th}$ station each consist of 3 layers single gap Resistive Plate Chambers (RPC), which will be operated at 15 kHz/cm$^2$ and 4 kHz/cm$^2$, respectively.

4. Physics performance studies

The performance studies discussed below are based on the configuration with the 1 m thick iron absorber as shown in the right panel of figure 3, which is designed for the measurement of the thermal radiation in the invariant mass range between $1 - 3$ MeV/$c^2$ and of J/$\psi$ mesons at beam energies above 4A GeV. The study was performed for central Au+Au collisions at a beam momentum of 8 A GeV/c. The background was generated by the UrQMD code, which describes hadronic interactions in terms of interactions between known hadrons and their resonances [12]. The embedded dimuon signals are calculated by the PLUTO generator which provides thermal particle distributions and collective flow [13]. The signal multiplicities are taken from the PHSD microscopic off-shell transport code [14]. The dimuon thermal radiation from the hot fireball was calculated with a coarse-grain approach [5]. The created particles are propagated through the detector systems using the GEANT3 transport code, which takes into account a realistic geometry and material.

The particle identification starts with the reconstruction of charged particle tracks in the STS using a Cellular Automaton algorithm, which is part of the CBM reconstruction software package. The reconstruction efficiency and the momentum resolution for charged particles emitted in central Au+Au collisions at 8A GeV/c are shown in figure 4. Then, the reconstructed tracks will be extrapolated through the detectors downstream the STS, i.e., the MuCh, TRD, and the TOF. The energy loss of the particles in the absorbers is calculated using the Bethe-Bloch equation, based on the muon mass hypothesis. The selection of muon candidates includes the following steps: cut on the primary collision vertex and pass all detector systems, cuts on the number of hits and the quality of the tracks in STS, MuCh and TRD, and cut on the particle mass based on momentum and time-of-flight as illustrated in figure 5. These cuts are optimized using a machine learning procedure.

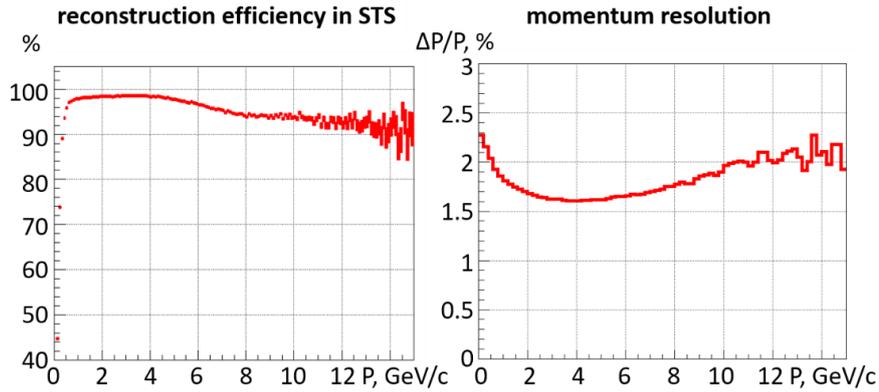

Fig. 4: Reconstruction efficiency for charged particles in the STS (left) and the corresponding momentum resolution (right) for central Au+Au collisions at 8A GeV/c.

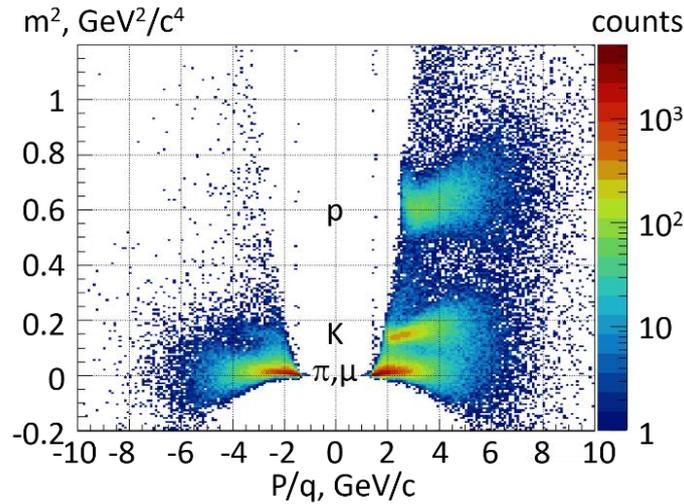

Fig. 5: Mass squared versus momentum times charge of particles reconstructed in central Au+Au collisions at 8A GeV/c.

The signal input into the simulations is illustrated in figure 6, which shows the µ⁺µ⁻ invariant mass distributions from the decays of the η meson and of the vector mesons up to the J/ψ meson. Moreover, the thermal radiation from the hot fireball is included [6]. The reconstructed invariant mass distribution is depicted in the left panel of figure 7 in the range from 0.2 GeV/$c^2$ to 2 GeV/$c^2$. The η and the φ mesons are clearly identified, and also an indication of the ω meson is seen. The resulting signal-to-background (S/B) ratio is shown in the right panel of figure 7. The average S/B value is in the order of 0.1 or better, which is an excellent value for dilepton measurements. In particular, the high S/B ratio at masses between 1 and 2 GeV/$c^2$ will allow to extract the thermal radiation from the fireball with high precision.

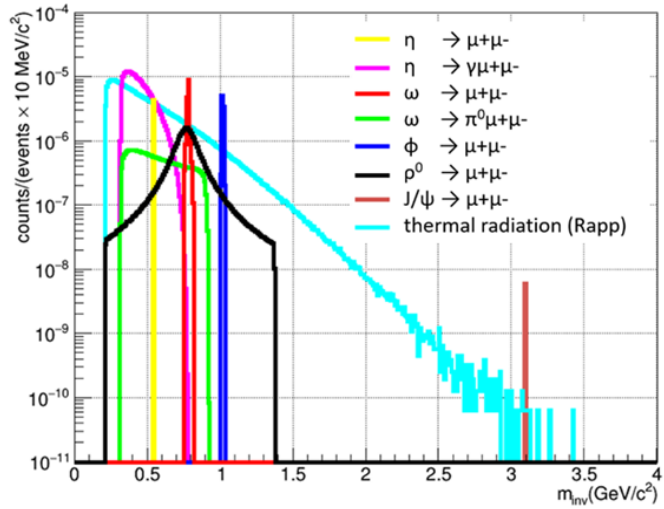

Fig. 6: Dimuon invariant mass distributions from the decays of η, ω, ρ, φ, and J/ψ mesons, and from thermal radiation [6] from the fireball in central Au+Au collisions at 8A GeV/c, as calculated with the PLUTO generator [13].

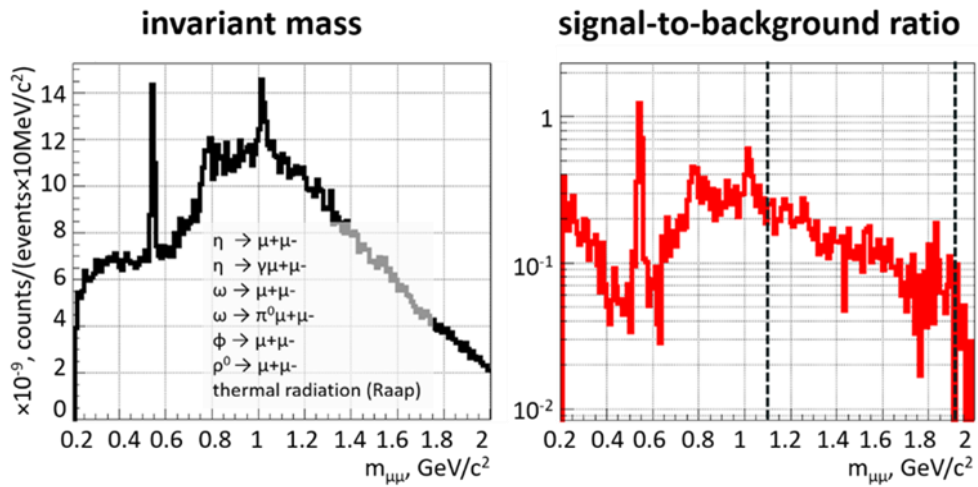

Fig. 7: Left panel: Reconstructed invariant mass distributions including the decays of η, ω, ρ, and φ mesons, and from thermal radiation from the fireball in central Au+Au collisions at 8A GeV/c. Right panel: Corresponding signal-to-background ratio.

The reconstructed dimuon invariant mass spectrum in the mass range between 2.8 and 3.4 GeV/c2 is depicted in the left panel of figure 8. The simulation was performed with the PLUTO code, taking into account the subthreshold J/ψ meson production mechanism via the decay of excited baryon resonances proposed by UrQMD [10]. J/ψ mesons are identified with a S/B ratio of 0.2 and an averaged efficiency of 2 %. The momentum dependent efficiency is shown in the right panel of figure 8. J/ψ mesons with momenta below about 6 GeV/c cannot be measured, because the J/ψ decay muons with momenta below about 2.6 GeV/c are absorbed.

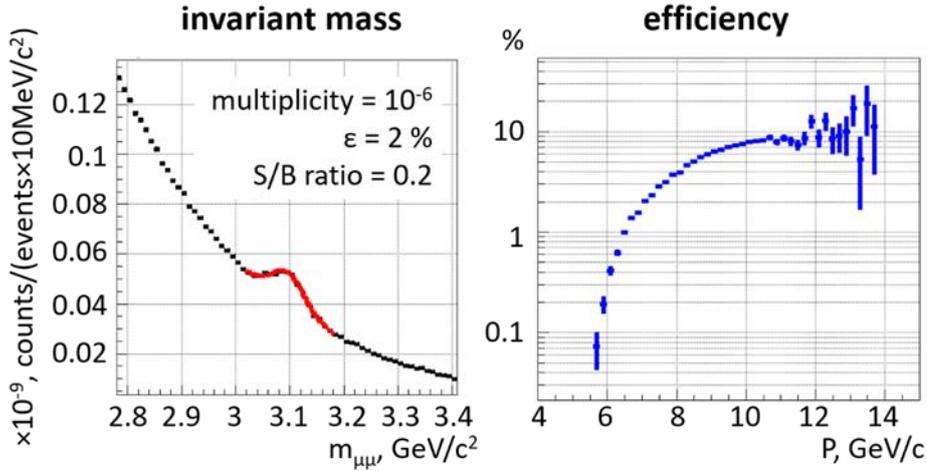

Fig. 8: Left: Reconstructed dimuon invariant mass spectrum calculated for central Au+Au collisions at 8A GeV/c using the PLUTO code including subthreshold J/ψ production [9] (see text). Right: efficiency of J/ψ meson reconstruction as function of J/ψ momentum.

The phase-space acceptance for J/ψ mesons is only little affected by the loss of the decay muons with low momenta. This is demonstrated in figure 9, which shows the simulation input as function of transverse momentum versus rapidity in the left panel, in comparison to the reconstructed J/ψ yield in the right panel. The distribution of reconstructed J/ψ mesons covers the same transverse momentum range, the acceptance is somewhat reduced at lower rapidity so that the maximum is slightly shifted towards larger rapidity, but still is close to mid rapidity.

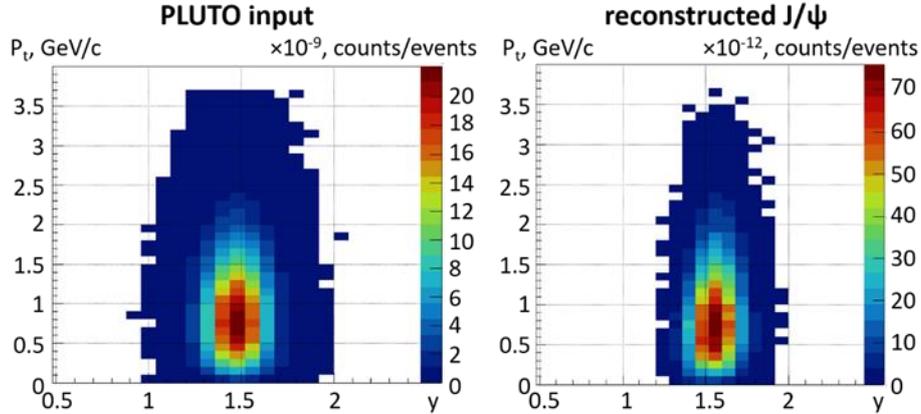

Fig. 9: Left: J/ψ meson simulation input in the plane transverse momentum versus rapidity for central Au+Au collisions at 8A GeV/c. Right: Corresponding reconstructed J/ψ meson distribution.

5.   Summary

The muon detection system of the CBM experiment at FAIR plays an important role for the execution of the envisaged research program, in particular for the exploration of the QCD phase diagram at large baryon chemicals. The invariant mass distribution of lepton pairs reflects in-medium modifications of ρ mesons, which might indicate chiral symmetry restoration, and the temperature of the fireball, which allows to search for the caloric curve of QCD matter. This opens the unique possibility, to discover or to rule out the existence of the critical endpoint of a 1$^{st}$ order phase transition, because the critical endpoint – if it exists - is located at temperatures below 130 MeV according to recent lattice QCD calculations. These temperatures are reached in the fireball created in heavy-ion collisions at FAIR energies. Moreover, the measurement of J/ψ mesons at subthreshold beam energies via μ$^+$μ$^-$ pairs will shed light on the degrees-of-freedom of QCD matter at high net-baryon densities. In conclusion, the dilepton measurements with the CBM experiment at FAIR represent an extremely promising and unrivaled research program with a high discovery potential concerning fundamental properties of QCD matter under extreme conditions.

Acknowledgement

The CBM project receives funding from the Europeans Union's Horizon 2020 research and innovation programme under grant agreement No. 871072. P. Senger acknowledges support from RFBR according to the research project No. 18-02-40086 by the Ministry of Science and Higher Education of the Russian Federation, Project "Fundamental properties of elementary particles and cosmology" No 0723-2020-0041.